\documentstyle[aps]{revtex}

\begin{document}

%%%%%%%%%%%%%%%%%%%%%%%%%%%%%%%%%%%%%%%%%%%%%%%%%%%%%%%%%%%%

\draft

\title{Generalized entropy and Noether charge}

\author{David Garfinkle \thanks{
Email: garfinkl@oakland.edu}}
\address{Department of Physics, Oakland University,
Rochester, Michigan 48309}
\author{Robert Mann \thanks{
Email: mann@avatar.UWaterloo.ca}}
\address{Department of Physics, University of Waterloo, Waterloo,
Ontario, N2L 3G1 Canada}

\maketitle

%%%%%%%%%%%%%%%%%%%%%%%%%%%%%%%%%%%%%%%%%%%%%%%%%%%%%%%%%%%%%%

\begin{abstract}

We find an expression for the generalized gravitational entropy of Hawking
in terms of Noether charge. As an example, the entropy of the Taub-Bolt
spacetime is calculated.

\end{abstract}

%%%%%%%%%%%%%%%%%%%%%%%%%%%%%%%%%%%%%%%%%%%%%%%%%%%%%%%%%%%%%%

\section{Introduction}

Gravitational thermodynamics is intimately and deeply connected with the
existence and properties of black holes \cite{beck,hawking0}. The area of
the event horizon of a black hole can be identified with 4 times its
physical entropy $S$, and its surface gravity is proportional to its
temperature $\beta ^{-1}$. These concepts can be extended to cosmological
horizons, where the same basic identifications come into play.

One way of understanding black hole entropy comes from the use of Euclidean
spacetimes. In the Euclidean analog of a black hole spacetime, the horizon
is a place where the Killing field, $\partial /\partial \tau $ vanishes and
is therefore an obstruction to foliating the spacetime with surfaces of
constant $\tau $. Such obstructions give rise to entropy by leading to a
difference between the Euclidean action $I$ and $H\Delta \tau $ where $H$ is
the Hamiltonian and $\Delta \tau $ is the period of the Euclidean time. More
generally, it has recently been pointed out that black holes are not the
only sources of gravitational entropy \cite{hawking1,hawking2}. In a $d$
dimensional Euclidean spacetime, a black hole horizon (also called a
``bolt'') is a fixed point set of the $U(1)$ isometry with dimension ${d_{f}}
=d-2$. However, such a spacetime may also contain a fixed point set with ${
d_{f}}<d-2$ (called a ``nut''). Associated with each nut is a set, called a
Misner string, where the $\tau $ foliation does not exist \cite{gibbons}. In
general a spacetime can contain both black holes and Misner strings, and the
total gravitational entropy will receive contributions from both of these
objects.

What then replaces the relation $S=A/4$ for this more general gravitational
entropy? For the general case, one has 
\begin{equation}
S=H\,\Delta \tau \,-\,I  \label{entropy}
\end{equation}
so one can simply find the entropy by calculating the energy and action.
However, this approach requires some care. In general, the action and
Hamiltonian are formally infinite, so one must introduce a background
spacetime and subtract off the background action (and Hamiltonian) to obtain
a finite answer. Alternatively, by adding to the action an additional
boundary term which is a functional of the intrinsic curvature invariants on
the boundary (so that the equations of motion are unaffected), the Misner
string contribution to the entropy can be intrinsically defined \cite{mann},
even if no bolts are present. The inclusion of this boundary term is
motivated from recent work \cite{bk} on the conjectured AdS/CFT duality,
which equates the bulk gravitational action of an asymptotically AdS
spacetime with the quantum effective action of a conformal field theory
(CFT) defined on the AdS boundary. The utility of such boundary terms
resides in their ability to render computation of the parameters of
gravitational thermodynamics finite without the inclusion of any reference
spacetime and the associated problematic issue of embedding. The
coefficients in the additional boundary terms may be uniquely fixed by
demanding such parameters are finite for, say, Schwarzchild-AdS spacetime.
However the spacetime need not be locally AdS asymptotically -- locally
asymptotically flat cases may also be included \cite{mann}.

Another approach to gravitational entropy is the Noether charge method of
Wald \cite{wald,waldiyer}. Here, the Noether charge is the integral 
of a $d-2$ form
associated with the diffeomorphism invariance of the theory. In this paper,
we address the question whether the generalized gravitational entropy of
Hawking can be expressed in terms of Noether charge. In section 2, we derive
a formula (equation (\ref{ncentropy}) below) relating generalized
gravitational entropy to Noether charge. In section 3 we use the Taub-Bolt
metric as an example of how to calculate gravitational entropy using both
equation (\ref{entropy}) and equation (\ref{ncentropy}).

\section{\protect\bigskip Noether Charge}

Consider the action 
\begin{equation}
I= - \, {\int_{M}}\;{\bf L}\;+\;{\int_{\partial M}}\;{\bf B}  \label{e1}
\end{equation}
for a manifold $M$ with boundary $\partial M$. We will consider the class of
metrics that approach some background metric ${\bar g}_{ab}$ at infinity and
will take the limit as $\partial M $ goes to infinity. For convenience we
will work with Euclidean manifolds throughout this paper. In General
Relativity with no matter fields, the Lagrangian ${\bf L}$ is given by 
\begin{equation}
{L_{abcd}}={\frac{1}{{16\pi }}}\;R\;{\epsilon _{abcd}}\;\;\;,  \label{e2}
\end{equation}
but more generally the action will be a functional of the metric and the
other matter fields. We denote the collection of these by $\Phi $.

As shown in reference\cite{wald}, there is a $d-2$ form ${\bf Q}$ associated
with the diffeomorphism invariance of the theory. This $d-2$ form arises as
follows: (for a more complete treatement see\cite{wald}). Since the action
is unchanged to first order by compact support variations of solutions to
the equations of motion, there is a $d-1$ form ${\bf \Theta}$ such that 
\begin{equation}
\delta {\bf L=E}\delta \Phi +d{\bf \Theta }(\Phi ,\delta \Phi )  \label{e3}
\end{equation}
Here the term ${\bf E}$ symbolically denotes the equations of motion. Since
the action of a solution $\Phi$ of the equations of motion must also be
stationary under variations that behave appropriately on the boundary $
\partial M$, it follows that 
\begin{eqnarray}
0 = \delta I &=& - \, {\int_{M}}\;\delta {\bf L}\;+\;{\int_{\partial M}}
\;\delta {\bf B=} - \; {\int_{M}}\;\left[ {\bf E}\delta \Phi +d{\bf \Theta }
(\Phi ,\delta \Phi )\right] \; + \;{\int_{\partial M}}\;\delta {\bf B} 
\nonumber \\
&=&{\int_{\partial M}}\;\left( \delta {\bf B} - {\bf \Theta }(\Phi ,\delta
\Phi )\right)  \label{e4}
\end{eqnarray}
Here the boundary term ${\bf B}$ is designed so that this equality is
satisfied. For general relativity with no matter fields one often takes ${
B_{abc}} = c K {\epsilon _{abc}}$ where $c$ is a constant and $K$ and $
\epsilon _{abc}$ are respectively the extrinsic curvature and the intrinsic
volume element of $\partial M$. However, this requires that one subtract off
the corresponding term in the background spacetime. Instead, we use the
following choice for ${\bf B}$. Let $D_a$ be the derivative operator of the
background metric ${\bar g}_{ab}$. Define ${\gamma _{ab}} = {g_{ab}} - {{
\bar g}_{ab}}$ and ${s_a} = {D^b} {\gamma _{ab}} - {D_a} \gamma$. Then ${\bf 
B}$ is given by 
\begin{equation}
{B_{abc}}={\frac{-1}{{16\pi }}}\;{\epsilon _{abcd}}{s^{d}}  \label{Bdef}
\end{equation}

There is a Noether current associated with every diffeomorphism generated by
a smooth vector field $\xi $. This current is an \ $(n-1)$-form ${\bf J}$
defined by 
\begin{equation}
{\bf J}\left[ \xi \right] {\bf =\Theta }(\Phi ,{{\cal L}_{\xi }}\Phi )-\xi
\cdot {\bf L}  \label{e5}
\end{equation}
where the ``dot'' denotes contraction on the first index of the form. \ It
is straightforward to show that 
\begin{equation}
d{{\bf J=-E}{{\cal L}_{\xi }}\Phi } =0  \label{e6}
\end{equation}
where the last equality holds whenever the equations of motion are
satisfied. \ Hence we can write 
\begin{equation}
{\bf J}\left[ \xi \right] =d{\bf Q}\left[ \xi \right]  \label{e7}
\end{equation}
where ${\bf Q}\left[ \xi \right] $ is the Noether charge \ $(n-2)$-form. In
general relativity ${Q_{ab}}[t]= - {\epsilon _{abcd}}{\nabla ^{c}}{t^{d}}
/(16\pi )$.

Consider the situation where $\partial M$ orthogonally intersects
hypersurfaces \ $\Sigma _{\tau}$ which foliate the manifold $M$ . The
intersections are labelled by compact $(n-2)$-surfaces $C_{\tau}$\ . The
Hamiltonian conjugate to a time-evolution vector field \ $t^{a}$\ (which
satisfies \ ${t^a}{\nabla _a}\tau =1$\ ) is defined to be a functional of
the fields and their derivatives whose variation is the integral of the
symplectic current 
\begin{eqnarray}
\delta H &=&{\int_{\Sigma _{\tau}}}\left[ \delta {\bf \Theta }(\Phi ,{{}_t}
\Phi )-{{\cal L}_t}{\bf \Theta }(\Phi ,\delta \Phi )\right]  \nonumber \\
&=&{\int_{\Sigma _{\tau}}}\left[ \delta {\bf J}\left[ t\right] -d\left(
t\cdot {\bf \Theta }(\Phi ,\delta \Phi )\right) \right]  \label{e8}
\end{eqnarray}
which implies 
\begin{equation}
H={\int_{\infty }}\;\left( {\bf Q[}t{\bf ]}\,-\,t\cdot {\bf B} \, + \, {\bf C
} \right)  \label{ham}
\end{equation}
where ${\bf C}$ is any quantity with zero variation.  Note that this
sort of foliation is incompatible with fixed points of $t^a$.  Thus, 
in order to define the Hamiltonian, we must remove from the manifold
any fixed points (and other obstructions to foliation). 
Since we want the
Hamiltonian to vanish in the background, we choose ${\bf C} = - {\bf \bar Q}$
where ${\bf \bar Q}$ denotes ${\bf Q}$ evaluated in the background. We
therefore have 
\begin{equation}
{H}={\int_{\infty }}\;\left( {\bf Q}\,-\,{\bar{{\bf Q}}}\,-\,t\cdot {\bf B}
\right)  \label{e10}
\end{equation}
One can show (see appendix) that this expression for the Hamiltionian is the
same as that of Abbott and Deser \cite{Abbdes}.

We now derive an expression for the entropy in terms of the Noether charge.
For a stationary solution, we remove (if necessary) a set of measure zero
from $M$ so that the resulting manifold can be foliated with surfaces of
constant $\tau $. The action is then given by 
\begin{equation}
I=-\Delta \tau \left( {\int_{\Sigma \tau }}\;t\cdot {\bf L}\;+\;{
\int_{\infty }}\;\,t\cdot {\bf B}\right)  \label{e11}
\end{equation}
Here the change in sign of the term involving ${\bf B}$ has to do with the
orientation of $\partial M$. Then using equation (\ref{entropy}) we find 
\begin{eqnarray}
S &=&\Delta \tau \,\left( {\int_{\infty }}\;\left( {\bf Q}\,-\,{\bar{{\bf Q}}
}\,-\,t\cdot {\bf B}\right) +{\int_{\Sigma \tau }}\;t\cdot {\bf L}\;+\;{
\int_{\infty }}\;\,t\cdot {\bf B}\right) \;\;\;  \nonumber \\
&=&\Delta \tau \left( {\int_{\infty }}\;{\bf Q}-{\int_{\Sigma \tau }}\;{\bf J
}\left[ t\right] \;\;-\;{\int_{\infty }}\;{\bar{{\bf Q}}}\right)
\end{eqnarray}
Since ${\bf J} = d {\bf Q}$, the integral of ${\bf J}$ depends only on ${\bf 
Q}$ at the boundary of $\Sigma _\tau$. Though $M$ has only $\partial M$ for
a boundary, we had to remove all the obstructions to foliation before
performing the integration. Let ${\cal O}$ be the intersection of these
obstructions with $\Sigma _\tau$. Then we have 
\begin{equation}
{\int_{\Sigma \tau }}\; {\bf J} \left[ t\right] = {\int_{\infty }}\; {\bf Q}
\; + \; {\int_{{\cal O}}}\; {\bf Q}
\end{equation}
We therefore find 
\begin{equation}
S = - \Delta \tau \left( {\int_{{\cal O}}}\;{\bf Q}\;+\;{\int_{\infty }}\;{
\bar{{\bf Q}}}\right)  \label{ncentropy}
\end{equation}
Thus, though Hawking's generalized entropy is no longer simply event horizon
area, it is Noether charge.

\section{Example}

The general form of the Taub-bolt/NUT metric is 
\begin{equation}
ds^{2}=V(r)(d\tau +2n\cos \theta d\phi )^{2}+{\frac{1}{V(r)}}
dr^{2}+(r^{2}-n^{2})(d\theta ^{2}+\sin ^{2}\theta d\phi ^{2})
\end{equation}
in Euclidean space, where 
\begin{equation}
V(r)=\frac{r^{2}-2mr+n^{2}}{r^{2}-n^{2}}
\end{equation}
and $m$ and $n$ are constants. Regularity in the Euclidean sector implies
that there be no string singularities, in which case the period of $\tau $
is $\Delta \tau =8\pi n$, whereas the absence of conical deficits in the $
(r,\tau )$ section implies the period of $\tau $ is $\Delta \tau =4\pi
/V^{\prime }({r_h})$ where $V({r_h})=0$. Requiring that these two periods be
commensurate yields 
\begin{equation}
m=\frac{5}{4}n\text{ \ \ \ \ \ \ \ \ \ \ \ \ \ or \ \ \ \ \ \ \ \ \ \ \ \ \
\ \ }m=n
\end{equation}
for which the event horizons are respectively at $r=2n$ (the bolt solution)
or $r=n$ (the NUT solution).

Here, we choose orientations so that $(\tau ,r,\theta ,\phi)$ is a right
handed coordinate system for spacetime and $(r,\theta ,\phi )$ is a right
handed coordinate system for the surfaces of constant $\tau$.

We now calculate the entropy of the Taub-Bolt metric, first using equation (
\ref{entropy}) and then using equation (\ref{ncentropy}). Here, the
background spacetime is the Taub-NUT metric. We begin by finding the action.
Since ${\bf L}$ vanishes in vacuum, the only contribution to the action
comes from ${\bf B}$. At large $r$ and to lowest order in ${r^{-1}}$ we have 
\begin{equation}
{B_{\tau \theta \phi }} = {\frac{{- n} }{{32 \pi}}} \; \sin \theta
\label{Bbolt}
\end{equation}
We therefore find 
\begin{equation}
I = {\frac{n }{8}} \; \Delta \tau = \pi {n^2}
\end{equation}
We now calculate the Hamiltonian. From equation (\ref{Bbolt}) it follows
that 
\begin{equation}
{\int _\infty } \; t \cdot {\bf B} = {\frac{{- n} }{8}}  \label{tdotBbolt}
\end{equation}
The components of ${\bf Q}$ that we will need are 
\begin{equation}
{Q_{\theta \phi }} = {\frac{1 }{{16 \pi }}} \; ( {r^2} - {n^2} ) \, {\frac{{
d V} }{{d r}}} \; \sin \theta  \label{Qthetaphi}
\end{equation}
\begin{equation}
{Q_{r \phi }} = {\frac{{n^2} }{{4 \pi }}} \; {\frac{V }{{{r^2} - {n^2}}}} \;
\cos \theta
\end{equation}
Using the form of the Taub-Bolt metric in equation (\ref{Qthetaphi}) it
follows that 
\begin{equation}
{\int _\infty} \; {\bf Q} = {\frac{{5 n} }{8}}  \label{Qbolt}
\end{equation}
The corresponding result using the Taub-Nut metric is 
\begin{equation}
{\int _\infty} \; {\bf \bar Q} = {\frac{n }{2}}  \label{Qnut}
\end{equation}
Now using the results of equations (\ref{tdotBbolt}, \ref{Qbolt}) and (\ref
{Qnut}) in equation (\ref{ham}) we find 
\begin{equation}
H = {\frac{n }{4}}
\end{equation}
Then using equation (\ref{entropy}) we find 
\begin{equation}
S = \pi {n^2}  \label{boltentropy}
\end{equation}
These results for $I, H $ and $S$ are all in agreement with those of
reference \cite{hawking1}. But note that no infinite background subtraction
was needed.

We now calculate the entropy using the Noether charge formula (equation (\ref
{ncentropy})). The obstructions to foliation are at $r =2 n, \; \theta = 0 $
and $\theta = \pi $. We then find 
\begin{eqnarray}
{\int _{{\cal O}}} \; {\bf Q} &=& {\lim _{\epsilon \to 0}} \; \left [ {\int
_{r = 2n +\epsilon}} \; {\bf Q} \; + {\int _{\theta = \epsilon}} \; {\bf Q}
\; + {\int _{\theta = \pi - \epsilon}} \; {\bf Q} \right ]  \nonumber \\
&=& {\lim _{\epsilon \to 0}} \; \left [ - \; {\int _0 ^{2 \pi}} \; d \phi \; 
{\int _0 ^\pi } \; d \theta \; {Q_{\theta \phi }} (2n+\epsilon , \theta ,
\phi ) \; - \; {\int _0 ^{2 \pi}} \; d \phi \; {\int _{2n} ^\infty } \; d r
\; {Q_{r \phi }} (r, \epsilon , \phi ) \; + \; {\int _0 ^{2 \pi}} \; d \phi
\; {\int _{2n} ^\infty } \; d r \; {Q_{r \phi }} (r, \pi - \epsilon , \phi )
\right ]  \nonumber \\
&=& - \; {\frac{{3 n} }{8}} \; - \; {\frac{n }{8}} \; - \; {\frac{n }{8}} =
- \; {\frac{{5 n} }{8}}  \label{Qobs}
\end{eqnarray}
Using equations (\ref{Qnut}) and (\ref{Qobs}) in equation (\ref{ncentropy})
we find $S = \pi {n^2} $ in agreement with equation (\ref{boltentropy}).
Thus the Noether charge method gives the same answer as the calculation of
entropy by first calculating the Hamiltonian and the action.

\section{Conclusions}

In its most general sense, gravitational entropy arises due
to an inability to everywhere foliate a given (Euclidean) 
spacetime with surfaces orthogonal
to a generator $\partial /\partial \tau $  of a $U(1)$ isometry. This
results in an inequality between the Hamiltonian and the action per isometry
period, the difference being them being the entropy. We have shown that
this form of entropy can be understood as the integral of the
Noether diffeomorphism charge integrated over all obstructions to
the foliation.  Our work is commensurate with recent work by Carlip \cite{Carlip} 
and Fatibene et. al. \cite{Fati}, who each arrived at similar conclusions
using different methods.

\section{Acknowledgements}

We would like to thank Bob Wald for helpful discussions. We would also like
to thank the Institute for Theoretical Physics at Santa Barbara (partially
supported by NSF grant PHY94-07194) for hospitality. DG was supported by NSF
grant PHY-9722039 to Oakland University and RM was supported by the Natural
Sciences and Engineering Research Council of Canada.

\section{Appendix}

Here, we show that the expression (\ref{e10}) for the Hamiltonian is
equivalent to that of Abbot and Deser \cite{Abbdes} which is 
\begin{equation}
{H_{AD}}={\frac{1}{{8\pi }}}\;{\int_{\infty }}\;{n_a}\,{u_c}\;\left[ {t_d}{
D_b}{K^{cadb}}\;-\;{K^{cbda }}\,{D_b}\,{t_d}\right]
\end{equation}
where ${K^{cadb}}\equiv {{\bar{g}}^{c \lbrack b }h^{d ]a }}\,-\,{{\bar{g}}
^{a \lbrack b }h^{d ]c }}$ and ${h_{ab}}\equiv {\gamma _{ab }}\,-\,{\frac{1}{
2}}\,\gamma {{\bar{g}}_{ab}}$. Here $u_a $ is the normal to the spacelike
hypersurface $\Sigma $ (where ${t_a}= - N{u_a}+{V_a}$) and $n_a$ is the
vector in $\Sigma $ that is normal to the two sphere at infinity. All
indicies are raised and lowered with the background metric. Using the
expression for $K^{cadb}$ and the antisymmetry of the Killing field
derivative we find 
\begin{equation}
{H_{AD}}={\frac{1}{{16\pi }}}\;{\int_{\infty }}\;{n_a}\,{u_c}\;\left[ {t_d}
\,2\,{D_b}{K^{cadb}}\;+\;\left( {{\bar{g}}^{cd}}\,{h^{ab}}\,+\,{{\bar{g}}
^{ab}}\,{h^{cd}}\right) \,{D_b}{t_d}\right] \;\;\;.
\end{equation}

Since the metric approaches the background at infinity, it follows that at
infinity ${\bf Q}-{\bar{{\bf Q}}}$ is given by the variation of ${\bf Q}$
where $\gamma _{ab}$ is used as the variation of the metric.  From the
expression for ${\bf Q}$ we then find 
\begin{eqnarray}
{\int_{\infty }}\;{\bf Q}\,-\,{\bar{{\bf Q}}} &=&{\frac{1}{{16\pi }}}\;{
\int_{\infty }}\;{n_a}\,{u_c}\;\left[ \left( {{\bar{g}}^{ab}}\,{h^{cd}}\,+\,{
{\bar{g}}^{cd}}\,{h^{ab }}\right) \,{D_b}{t_d}\;\right.  \nonumber \\
&&\left. +\;{t_d}\,{D_b}\left( {{\bar{g}}^{cb}}\,{h^{ad}}\,-\;{{\bar{g}}^{ab}
}\,{h^{cd}}\,-\,{\frac{1}{2}}{{\bar{g}}^{da}}\,{{\bar{g}}^{cb}}\,h\,+\,{
\frac{1}{2}}{{\bar{g}}^{dc}}\,{{\bar{g}}^{ab}}\,h\right) \right] \;\;\;.
\end{eqnarray}

\bigskip

Now, comparing this expression to the one for $H_{AD}$ gives 
\begin{eqnarray}
{H_{AD}} &=&{\int_{\infty }}\;{\bf Q}\, - \,{\bar{{\bf Q}}}\,  \nonumber \\
&&\,+\;{\frac{1}{{16\pi }}}\;{\int_{\infty }}\;{n_a}\,{u_c}\,{t_d}\,{D_b}
\;\left( 2{K^{cadb}}\,-\,{{\bar{g}}^{cb}}\,{h^{ad}}\,+\,{{\bar{g}}^{ab}}\,{
h^{c d}}\,+\,{\frac{1}{2}}\,h\,({{\bar{g}}^{da}}\,{{\bar{g}}^{c b}}\,-\,{{
\bar{g}}^{dc}}\,{{\bar{g}}^{ab}})\right) \;\;\;.
\end{eqnarray}
Then using the form of ${K^{cadb}}$ and $H_{ab}$ we find 
\begin{eqnarray}
{H_{AD}} &=&{\int_\infty }\;{\bf Q}\, - \,{\bar{\bf Q}} 
+\;{\frac{1}{{16\pi }}}\;{
\int_\infty }{n_a}{u_c}{t_d}\,{D_b}\;\left( {{\bar{g}}^{ad}}\,({\gamma ^{bc}}
\,-\,\gamma {{\bar{g}}^{bd}})\,-\,{{\bar g}^{cd}}\,({\gamma ^{ab}}
\,-\,\gamma \,{{\bar g}^{ab}})\right) \;\;\;.  \nonumber \\
&=&{\int_\infty }\;{\bf Q}\,-\,{\bar{\bf Q}} 
+\;{\frac{1}{{16\pi }}}\;{\int_{\infty }}
\;{n_a}\,{u_c}\,{t_d}\,({{\bar{g}}^{ad}}\,{s^c}\,-\,{{\bar{g}}^{cd}}\,{s^a}
)\;
\end{eqnarray}
where the last line follows from the definition of $s_{a}$.

Then from the definition of ${\bf B}$ (equation (\ref{Bdef})) we obtain our
result: 
\begin{equation}
{H_{AD}}={\int_{\infty }}\;{\bf Q}\,-\,{\bar{{\bf Q}}}\,-\,t\cdot {\bf B}
\end{equation}


\bigskip

\begin{references}
\bibitem{beck}  J.D. Beckenstein, Phys. Rev. {\bf D7}, 2333 (1973)

\bibitem{hawking0}  S.W. Hawking, Nature {\bf 248}, 30 (1974); Commun. Math.
Phys. {\bf 43}, 199 (1975)

\bibitem{hawking1}  S.W. Hawking and C.J. Hunter, Phys. Rev. {\bf D59},
044025 (1999)

\bibitem{hawking2}  S.W. Hawking, C.J. Hunter and D.N. Page, Phys. Rev. {\bf 
D59}, 044033 (1999)

\bibitem{gibbons}  G.W. Gibbons and S.W. Hawking, Commun. Math. Phys. {\bf 66
}, 291 (1979)

\bibitem{mann}  R.B. Mann, Phys. Rev. {\bf D60}, 104047 (1999); R.B. Mann
Phys. Rev. {\bf D61}, 081043 (2000)

\bibitem{bk}  V. Balasubramanian and P. Kraus, Commun. Math. Phys. {\bf 208}
, 413 (1999)

\bibitem{wald}  R.M. Wald, Phys. Rev. {\bf D48}, R3427 (1993)

\bibitem{waldiyer}  V. Iyer and R.M. Wald, Phys. Rev. {\bf D52},
4430 (1995)
 
\bibitem{Abbdes}  L.F. Abbott and S. Deser, Nuclear Physics B {\bf 195}, 76
(1982)

\bibitem{Carlip}S. Carlip, Class. Quant. Grav. {\bf 16}, 3327 (1999).

\bibitem{Fati}L. Fatibene, M. Ferraris, M. Francaviglia and M. Raiteri,
{gr-qc/9906114}

\end{references}
\end{document}